\documentclass[preprint2]{aastex}

\usepackage{amsmath}
\usepackage{graphicx}

\newcommand{\snrg}{S/N($g$)}
\newcommand{\teff}{$T_{\rm eff}$}
\newcommand{\teffLM}{$T_{\rm eff,LM}$}
\newcommand{\logg}{log$g$}
\newcommand{\feh}{[Fe/H]}
\newcommand{\LM}{$_{\rm LM}$}
\newcommand{\sei}{$_{\rm ast}$}
\newcommand{\SVR}{$_{\rm SVR}$}

\slugcomment{}

\shorttitle{Asteroseismic based estimation of the surface gravity}
\shortauthors{Liu et al.}

\begin{document}


\title{Asteroseismic based estimation of the surface gravity for the LAMOST giant stars}


\author{Chao Liu\altaffilmark{1}, Min Fang\altaffilmark{2,3}, Yue Wu\altaffilmark{1}, Li-Cai Deng\altaffilmark{1}, Liang Wang\altaffilmark{1}, Wei Wang\altaffilmark{1}, Jian-Ning Fu\altaffilmark{4}, Yong-Hui Hou\altaffilmark{5}, Guang-Wei Li\altaffilmark{1}, Yong Zhang\altaffilmark{5}}
\email{liuchao@nao.cas.cn}


\altaffiltext{1}{Key Laboratory of Optical Astronomy, National Astronomical Observatories, Chinese Academy of Sciences, 20A Datun Road, Beijing 100012, China}
\altaffiltext{2}{Departamento de F\'isica Te\'orica, Facultad de Ciencias, Universidad Auton\'oma de Madrid, 28049 Cantoblanco, Madrid, Spain}
\altaffiltext{3}{Purple Mountain Observatory and Key Laboratory for Radio Astronomy, 2 West Beijing Road, 210008, Nanjing, China}
\altaffiltext{4}{Department of Astronomy, Beijing Normal University, 19 Avenue Xinjiekouwai, Beijing 100875, China}
\altaffiltext{5}{Nanjing Institute of Astronomical Optics \& Technology, National Astronomical Observatories, Chinese Academy of Sciences, Nanjing 210042, China}

\begin{abstract}

Asteroseismology is one of the most accurate approaches to estimate the surface gravity of a star. However, most of the data from the current spectroscopic surveys do not have asteroseismic measurements, which is very expensive and time consuming. In order to improve the spectroscopic surface gravity estimates for a large amount of survey data with the help of the small subset of the data with seismic measurements, we set up a support vector regression model for the estimation of the surface gravity supervised by  1,374 LAMOST giant stars with \emph{Kepler} seismic surface gravity. The new approach can reduce the uncertainty of the estimates down to about 0.1\,dex, which is better than {the LAMOST pipeline} by at least a factor of 2, for the spectra with signal-to-noise ratio higher than 20. {Compared with the \logg\ estimated from the LAMOST pipeline, the revised \logg\ values provide a significantly improved match to the expected distribution of red clump and RGB stars from stellar isochrones.} Moreover, even the red bump stars, which extend to only about 0.1\,dex in \logg, can be discriminated from the new estimated surface gravity. The method is then applied to about  {350,000} LAMOST metal-rich giant stars to provide improved surface gravity estimates.  {In general, the uncertainty of the distance estimate based on the SVR surface gravity can be reduced to about 12\% for the LAMOST data. 
}

\end{abstract}



\section{Introduction}

The surface gravity of a star is an important stellar astrophysical parameter in the sense that it is able to measure the radius of a star given the stellar mass. Together with the effective temperature and metallicity, a star can be pinned down in the Hertzsprung-Russell diagram with the surface gravity. This will be very helpful to learn the evolution status of the star as well as its distance. Therefore, an accurate estimation of the surface gravity is critical in the study of either stars or stellar systems. 

Although multi-band photometry may help to discriminate giant from dwarf stars according to some surface gravity sensitive features \citep{lenz98, majewski00, yanny00}, spectroscopic data can reveal more detailed features to quantify the surface gravity. First, the prominent Mg$b$+MgH feature observed in low resolution spectra is not only be used to identify giant stars \citep{xue14,liu14}, but also to measure their surface gravity \citep{morrison03,lee08}. Second, the other features, e.g., Balmer lines \citep{wilhelm99}, CaII K and H lines \citep{lee08} etc., can also be useful to the determination of the surface gravity. Moreover, some algorithms determine the surface gravity together with the effective temperature and metallicity, simultaneously, by comparing the full spectra with the spectral library \citep[e.g., ][]{lee08, wu11a}. In addition, the supervised machine learning approaches, e.g. artificial neural networks, support vector machine etc., have also been used to derive the surface gravity based on the training spectra with known surface gravity values as the targets \citep{refiorentin07,liu12}. The typical accuracy of the surface gravity estimates for low resolution spectra, e.g. SDSS \citep{ahn14} or LAMOST \citep{cui12}, is about 0.2-0.4\,dex \citep{wilhelm99, refiorentin07, lee08, wu14}.

Asteroseismology is a powerful tool to derive the fundamental parameters, e.g. stellar mass, radius, and {\logg}, for a star \citep[see][]{brown94,chaplin13}. Thanks to the \emph{Kepler} \citep{borucki10} mission, the solar-like oscillations for tens of thousands of stars are able to be measured. The surface gravity can be estimated from the oscillations with accuracy of 0.02$\sim$0.05\,dex \citep{morel12, creevey13}. This performance is much better than the non-seismic methods from even the high-resolution spectra. Indeed, \citet{epstein14b} has shown that the \emph{Kepler} measured seismic \logg\ is more accurate than those from the high-resolution infrared APOGEE \citep{majewski2010} spectra by a factor of a few.

However, compared to the huge amount of spectra from many large spectroscopic survey projects, the number of stars with asteroseismic measurement is still very limited. Therefore, it is very crucial to examine how the surface gravity of the whole spectroscopic survey data can be improved with the existing seismic data, which only occupies a small fraction of the full samples. In this  {paper}, we give more accurate \logg\ estimates for the LAMOST data with the help of a small subset of the spectra with \emph{Kepler} seismic \logg.

The LAMOST telescope, also known as Guoshoujing Telescope, is a new type of 5-degree wide field telescope with a large aperture of 4 meter. It assembles 4000 fibers on its large focal plane and can simultaneously observe the similar number of low-resolution ($R\sim1800$) spectra covering the wavelength from 380 to 900\,nm \citep{cui12,zhao12}. As its main scientific goal, it will observe a few millions of stellar spectra with limiting magnitude down to $r\sim18$\,mag for diverse studies of the Milky Way \citep{deng12}. It is also sampled the \emph{Kepler} field with the ``LAMOST-\emph{Kepler} project'' \citep{decat14} {, in which} a few thousands of spectra with \emph{Kepler} seismic data \citep{huber14} { are included}. This small subset provides perfect calibrators to improve the estimation of \logg\ for the LAMOST spectra. 

 {In this work, w}e develop a support vector regression  {(hereafter, SVR)} model for the determination of the surface gravity for the LAMOST giant stars supervised by the data with \emph{Kepler} seismic \logg.
 {The paper is organized as follows. In section~\ref{sect:svm}, we give a detailed introduction to SVR. Then, we specifically set up a SVR model to estimate \logg\ for LAMOST data in section~\ref{sect:result}. The performance of the determination is also assessed in this session. Some discussions, including the systematics from the seismic \logg, the influences in various evolution phases of the stars, the metallicity effect, the comparison with other similar works, and the benefits and limits of this technique, are raised in section~\ref{sect:discussions}. Finally, a brief conclusion is drawn in the final section.}

\section{Support vector regression}\label{sect:svm}
\begin{figure}[htbp]
\centering
\includegraphics[scale=0.7]{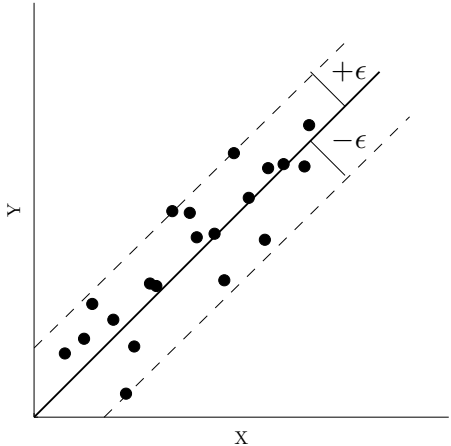}
\caption{ {The plot demonstrates how SVR model works in linear case. The black dots are the mock data with arbitrary scales in x- and y-axes. The solid line stands for the SVR model and the dashed lines show the tolerance of $\epsilon$ in each side of the linear model.}}\label{fig:SVR}
\end{figure}
 {Support vector machine (SVM) is a well known supervised machine learning algorithm mostly in the application of classification and nonlinear regression \citep{cortes95,burges98,deng12svm}.  A {SVR} \citep[][]{drucker96}, as an extension of the SVM, is a regression method by transforming the data, via a kernel function, from the non-linear physical space into a high dimensional inner-product space, in which a linear model to the data can be fitted. }

 {The complete description of SVR is referred to \citet{smola04}. We only give an outline of the approach. Assume that $\{(\mathbf{x}_1,{y}_1),$ $(\mathbf{x}_2,{y}_2),$ $\cdots,$ $(\mathbf{x}_\ell,{y}_\ell)\}$ are the training data, where $\mathbf{x}_i$ is the input vector in $d$ dimensions and ${y}_i$ is the corresponding known output. Because the relationship between $x$ and $y$ is highly non-linear, the SVR is to find a map, $\rm\Phi$, mapping the real input space to a higher dimensional feature space, in which the non-linear regression problem can be converted into a linear problem. For this purpose, the SVR constructs a form such as
\begin{equation}\label{eq:svrmodel1}
f(\mathbf{x})=\sum_{i}^{\ell}\langle(\alpha_i-\alpha_i^{\ast}){\rm\Phi(\mathbf{x}_i)},{\rm\Phi(\mathbf{x})\rangle}+b,
\end{equation}
where $\langle\cdot,\cdot\rangle$ is inner product, $\alpha_i$, $\alpha_i^{\ast}$, and $b$ are parameters to be solved. Practically, $\rm\Phi$ does not explicitly given, but use a kernel function, $K(\mathbf{x},\mathbf{x'})\equiv\langle\rm\Phi(x),\rm\Phi(x')\rangle$, instead. Among lots of choices of the kernels, the radial-based kernel, which has the form of $\mathrm{exp}(-\gamma\|\mathbf{x}-\mathbf{x'}\|)$, is often used. Mathematically, it can be proved that $f$ is linear in the inner product space. The solution of Equation (\ref{eq:svrmodel1}) can be derived from a convex optimization problem:
\begin{eqnarray}\label{eq:svrmodel2}
&\textrm{minimize }&{1\over{2}}\|\textbf w\|^2+C\sum_{i=1}^{\ell}(\xi_i+\xi_i^{\ast})\nonumber\\
&\textrm{subject to }&\left\{\begin{array}{l}y_i-\langle \textbf w,{\rm\Phi}(\textbf{x}_i)\rangle-b\leqslant\epsilon+\xi_i\\\langle \textbf w,{\rm\Phi}(\textbf{x}_i)\rangle+b-y_i\leqslant\epsilon+\xi_i^{\ast}\\
\xi_i,\xi_i^\ast\geqslant0\end{array}\right.
\end{eqnarray}
where $w=\sum_{i=1}^\ell{(\alpha_i-\alpha_i^{\ast}){\rm\Phi}(\mathbf{x}_i)}$, $C>0$ is the cost parameter to determine the trade-off between the flatness of $f$ and the amount up to which deviations larger than $\epsilon$ are tolerated, $\xi_i$ and $\xi_i^\ast$ are slack variables. Figure~\ref{fig:SVR} demonstrates how the SVR works in linear case. }

 {\citet{libsvm} provides a multi-programming language packages, LIBSVM\footnote{http://www.csie.ntu.edu.tw/$\sim$cjlin/libsvm/}, to solve the SVR model based on Equations (\ref{eq:svrmodel1}) and (\ref{eq:svrmodel2}) as well as other SVM models. The process of finding the optimized solution of the SVR model contains two steps. The first step is to determine the parameters $C$, $\epsilon$, and $\gamma$ from a $n$-fold cross-validations. In this step, the software randomly divides the training dataset into $n$ groups with equal size and predict the dependent values for each group using the SVR model trained by the rest $n$-1 groups of data. The best choices of $C$, $\epsilon$, and $\gamma$ are those who give the best cross-validation accuracy of the regression.  The second step is to find the optimal solution of $\alpha_i$, $\alpha_i^\ast$, $\xi_i$, $\xi_i^\ast$, and $b$ in Equation (\ref{eq:svrmodel2}) for the whole training dataset using the best choices of $C$, $\epsilon$, and $\gamma$. Then, the derived SVR model, Equation (\ref{eq:svrmodel1}), can be used to predict the corresponding dependent variable for a given input data. A typical application of the SVR is also found in \citet{liu12}.}

\section{Estimate \logg\ for LAMOST spectra}\label{sect:result}
\subsection{The training and test dataset}\label{sect:data}
\begin{figure}[htbp]
\centering
\includegraphics[scale=0.6]{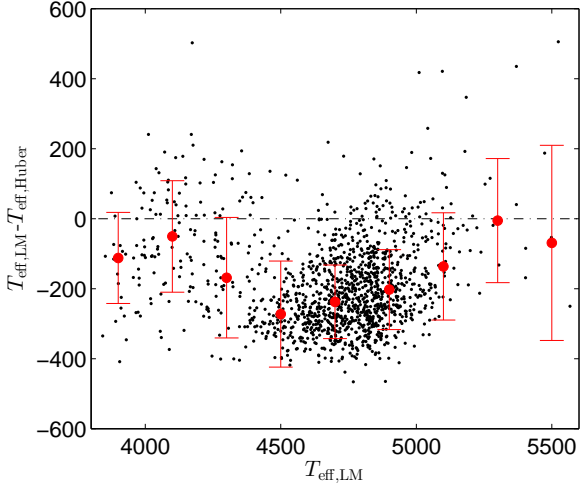}
\caption{ {The difference between the LAMOST \teff\ and the \teff\ in Huber catalog. The x-axis is the LAMOST \teff\ and the y-axis is the difference of the \teff\ between LAMOST and Huber catalog. The black dots stand for the individual stars in the training dataset. And the red filled circles with error bars indicate the medians and dispersions of the differences at various \teff\ bins.}}\label{fig:teffsystem}
\end{figure}
\begin{figure}[htbp]
\centering
\includegraphics[scale=0.6]{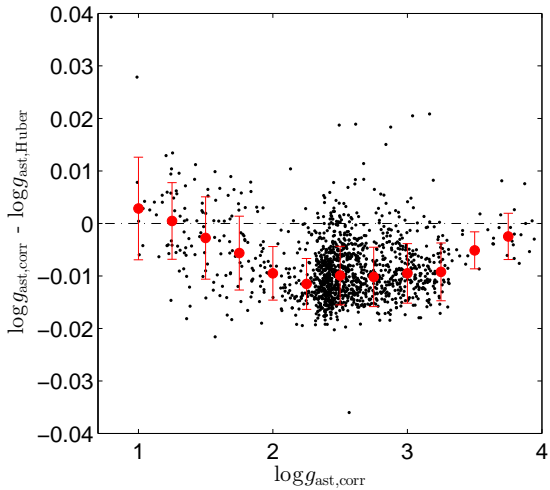}
\caption{ {The difference between the initial \logg\ from Huber catalog and the re-calibrated one using LAMOST \teffLM. The x-axis is the re-calibrated \logg, while the y-axis is the difference. The black dots stand for the stars in the training dataset. And the red filled circles with error bars are the medians and dispersions of the differences at various \logg$_{\rm ast,corr}$ bins.}}\label{fig:corrlogg}
\end{figure}
In order to establish a SVR model for the \logg\ determination based on the seismic estimates, we firstly select a proper training dataset.
\citet{huber14} released $\sim200,000$ stars with stellar parameters in the \emph{Kepler} field, among them, 15,686 stars have surface gravity estimated from asteroseismology  {hereafter, we call it as \emph{Huber catalog})}.  {We use the parameters listed in the table 4 of~\citet{huber14}, which is directly compiled from literatures and not fitted to the model isochrones.}  We cross-identify these stars with the LAMOST DR2 data and obtain  {3,335} common stars with the signal-to-noise ratio at $g$ band higher than 10 in the spectra\footnote{We use the signal-to-noise ratio at $g$ band because most of the spectral lines sensitive to \logg\ are located in this range of wavelength \citep{liu14}}. The stellar parameters of these LAMOST spectra are estimated in the pipeline using the software \emph{ULySS}, with the empirical stellar library ELODIE as a reference \citep{wu11a,wu11b,wu14}. 

 {The scaling relation of the seismic \logg, as shown in equation~(\ref{eq:scleqn}), depends on \teff.
\begin{equation}\label{eq:scleqn}
{g\over{{g}_\odot}}\cong\large({\nu_{max}\over{\nu_{max,\odot}}}\large)\large({T_{\rm eff}\over{T_{\rm{eff},\odot}}}\large)^{0.5},
\end{equation}
where $\nu_{max}$ is the frequency of maximum power and the quantities with $\cdot_{\odot}$ stand for the solar parameters~\citep{chaplin13}.
The effective temperature used to derive the seismic \logg\ in Huber catalog may not be consistent with the LAMOST \teff. Indeed, figure~\ref{fig:teffsystem} shows that the LAMOST \teff\ is smaller than those in Huber catalog by about 200\,K. Therefore, the initial seismic \logg\ in Huber catalog needs to be re-calibrated with LAMOST \teff. Without providing the frequencies of the seismic oscillation in Huber catalog, we can re-calibrate the seismic \logg\ be applying the following rescaling:
\begin{equation}\label{eq:caliblogg}
g_{\rm ast,corr}=g_{\rm ast, Huber}\large({T_{\rm eff,LM}\over{T_{\rm eff,Huber}}}\large)^{0.5},
\end{equation}
where $T_{\rm eff,LM}$ is the LAMOST derived effective temperature, $T_{\rm eff,Huber}$ and \logg$_{\rm ast,Huber}$ the effective temperature and the seismic surface gravity provided by Huber catalog, respectively. In order to keep self-consistence of \teff\ in Huber catalog, we only use the objects with~\citet{pinsonneault12} \teff\ or whose \teff\ is calibrated with~\citet{pinsonneault12} in Huber catalog. Consequently, a very small fraction of stars, which are from the reference 6, 13, and 14 and hence may not explicitly calibrate to ~\citet{pinsonneault12},  in Huber catalog are excluded from our training and test dataset.}

 { Figure~\ref{fig:corrlogg} shows the difference between the initial and the re-calibrated \logg. The re-calibrated \logg, denoted as \logg$_{\rm ast,corr}$, is slightly smaller than the initial \logg, denoted as \logg$_{\rm ast,Huber}$, by only about 0.01\,dex, which is essentially within the uncertainty of the seismic \logg. This trend does naturally follow the difference in \teff\ between the two catalogs. Although this tiny difference would not significantly affect our result, we still adopt the re-calibrated \logg\ in this work to keep well self-consistency between the LAMOST \teff\ and the seismic \logg. In the rest of the paper, unless explicitly indicated, we refer to the re-calibrated seismic \logg\ by default, when we use \logg\sei.} 

\begin{figure*}[htbp]
\centering
\includegraphics[scale=0.75]{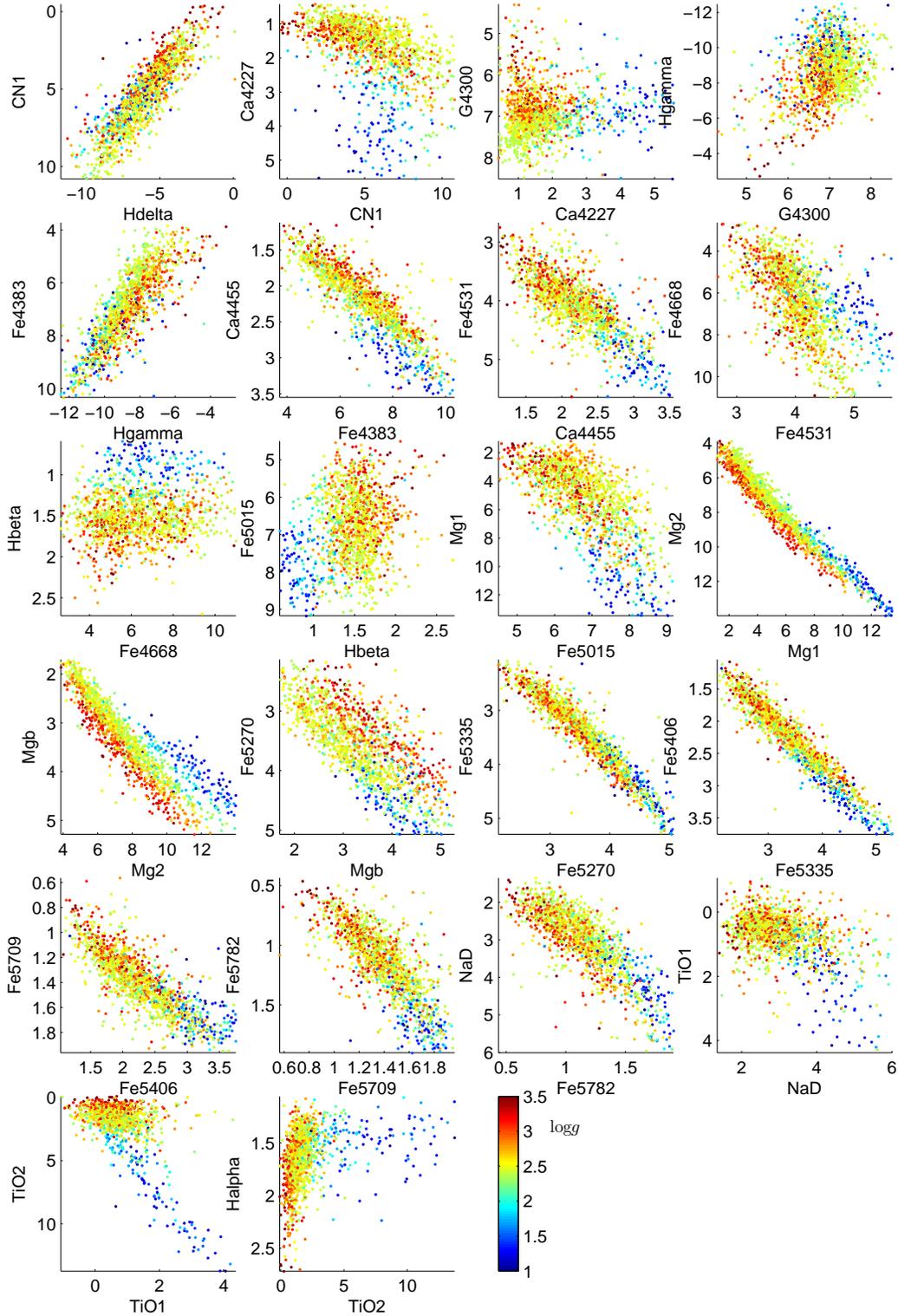}
\caption{ {The figure indicates the correlations between the \logg\sei\ and the 23 Lick indices for the training dataset. The x and y axes are two different Lick indices and the color codes the corrected seismic surface gravity.}}\label{fig:EWlogg}
\end{figure*}

\begin{figure*}[htbp]
\centering
\includegraphics[scale=0.75]{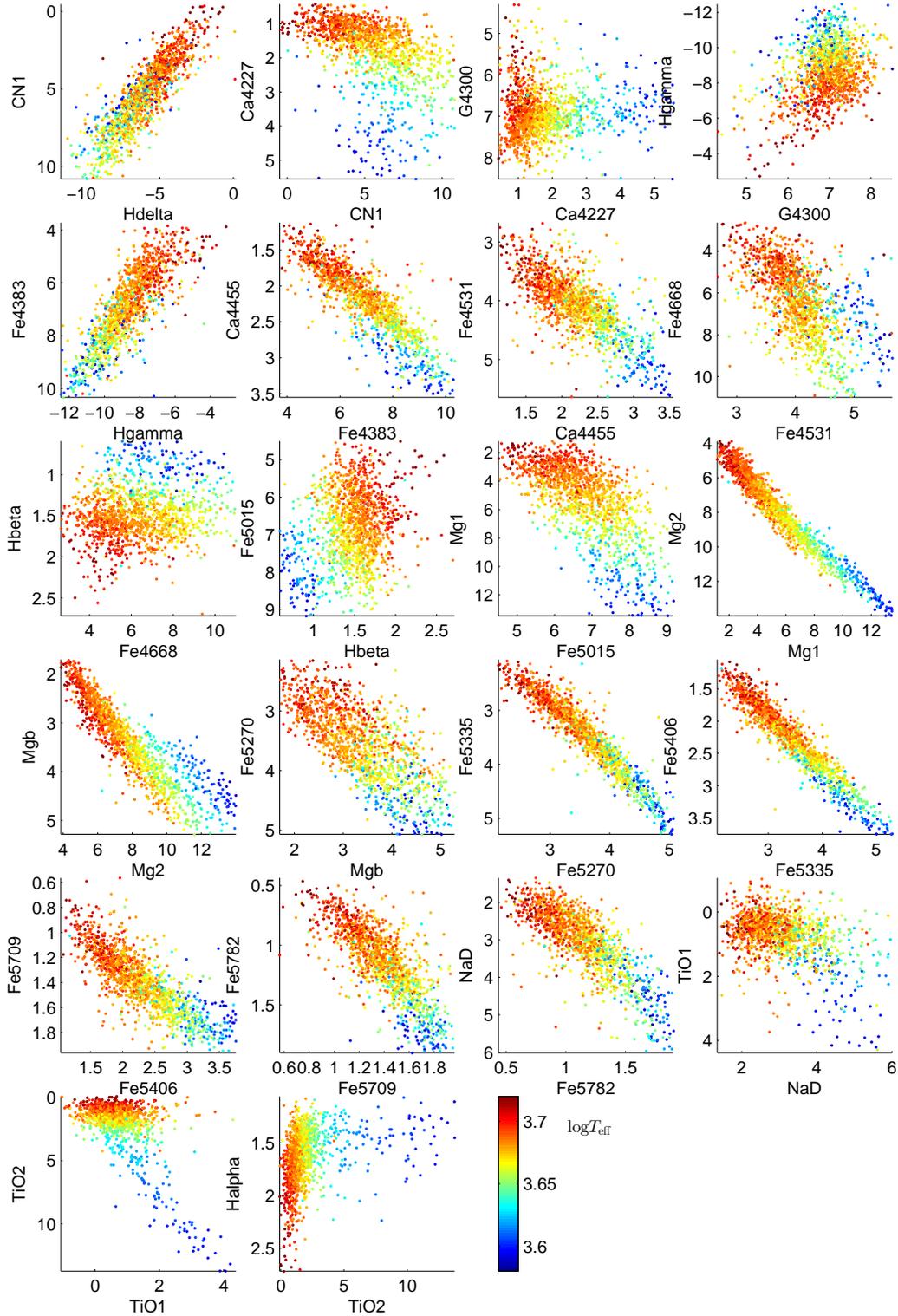}
\caption{ {The figure indicates the correlations between the \teff\ and the 23 Lick indices for the training dataset. The x and y axes are two different Lick indices and the color codes the effective temperature from LAMOST pipeline.}}\label{fig:EWteff}
\end{figure*}

\begin{figure*}[htbp]
\centering
\includegraphics[scale=0.75]{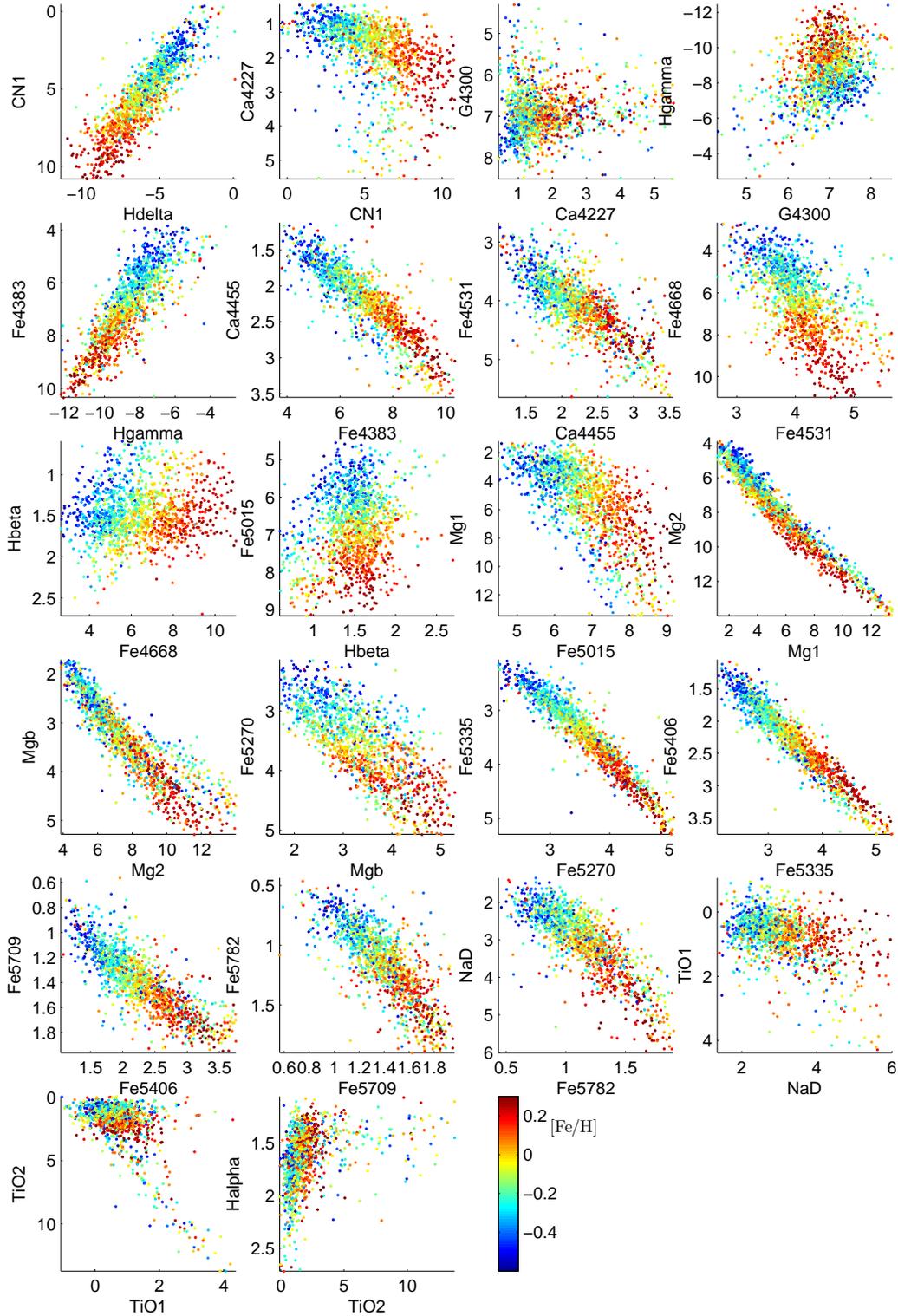}
\caption{ {The figure indicates the correlations between the \feh\ and the 23 Lick indices for the training dataset. The x and y axes are two different Lick indices and the color codes the metallicity from the LAMOST pipeline.}}\label{fig:EWfeh}
\end{figure*}

Then, we select the metal-rich giant stars with  {${3500<T_{\rm{effLM}}<6000}$\,K,} \logg\LM$<4.0$\,dex and \feh\LM$>-0.6$\,dex, where the subscript ``LM'' denotes that the parameters are from the LAMOST pipeline. The selection in metallicity is because that the current scaling relation in asteroseismology may not be suitable for the metal-poor stars, e.g., the thick disk or the halo stars \citep{epstein14a}. {The selection in loggLM is motivated by the small amount of seismic detections for dwarf stars in the Kepler database.} Finally, we select  {2,726} stars with both seismic \logg\ and LAMOST spectra with \emph{ULySS} derived \teffLM, \feh\LM, and \logg\LM. 

First, the data is arbitrarily separated into two groups with essentially equal members. One,  {known as the training dataset (1,374 stars),} is used to train the SVR to find the best fit model, the other,  {including the rest 1,352 stars}, is used as the test dataset to assess the performance.

Second, we define the proper quantities as inputs of the SVR model. We use the equivalent widths of 23 Lick lines \citep{worthey94, worthey97}, including H$_\delta$, H$_\gamma$, H$_\beta$, H$_\alpha$, Mg$_1$, Mg$_2$, Mg$_b$, Fe (4383, 4531, 4668, 5015, 5270, 5335, 5406, 5709, 5782$\rm\AA$), CN, Ca (4227, 4455$\rm\AA$), G band, NaD,  {and} TiO(5950, 6187-6269 $\rm\AA$), measured from the LAMOST spectra, rather than the full spectra, as the input to the SVR model.  {These line indices are more robust to the noise than the full spectrum.}

 {In general, if the line indices varies with \logg\ in the same way as with \teff, then it is not easy to distinguish whether the change of the line indices is because the change of \logg\ or \teff. This is so called degeneracy between \logg\ and \teff. For instance, H$_\alpha$, H$_\beta$, and TiO show similar trend of variation when either \logg\ or \teff\ changes (see Figures~\ref{fig:EWlogg} and~\ref{fig:EWteff}). However, not all of the 23 Lick line indices are affected by the degeneracy. The lines such as Mg$_1$, Mg$_2$, Mg$_b$, Fe5335, Fe5782 and etc. vary with \logg\ in a different way with \teff\ (also see Figures~\ref{fig:EWlogg} and~\ref{fig:EWteff}). Therefore, these lines are more helpful to break the degeneracy and allow us to estimate \logg. Figure~\ref{fig:EWfeh} also shows that the variations of \feh\ in the line indices are quite different with \logg, meaning that there is {relatively little degeneracy} between \feh\ and \logg.} 

 {Although the SVR model can only deal with one dimensional dependent variable, it may not produce systematically biased \logg\ estimates related to either \teff\ and \feh, because the training dataset covers the same ranges of the metallicity and effective temperature as the test dataset and the full LAMOST sample. More investigations will be discussed in section~\ref{sect:performance}.}

 {It is also worthy to notice that although the asteroseismic \logg\sei\ is used as the known dependent variable in the training dataset, it does not mean that the seismic \logg\ is the \emph{true} value for a star. It is also affected by some systematics, as mentioned later in section~\ref{sect:seismicbias}. However, compared to the relatively larger uncertainty of the spectroscopic \logg\ estimates, the systematic bias in the seismic \logg\ is quite small and can be negligible. Moreover, the motivation of this work is not to find the \emph{true} \logg\ for the spectra, but to find the better \logg\ estimates than the other spectroscopic methods. Since the seismic \logg\ is so far the best choice, the goal of this work is to calibrate the \logg\ estimated from the low-resolution spectra to the best one, i.e., the seismic \logg.} 

 {Note that the LAMOST pipeline derived \logg\LM\ is only used in the initial data selection; it plays no role in the SVR model. In other word, the SVR predicted \logg\SVR\ is independent with \logg\LM.}

\subsection{Performance}\label{sect:performance}
\begin{figure*}[ht]
\begin{center}
\includegraphics[scale=0.5]{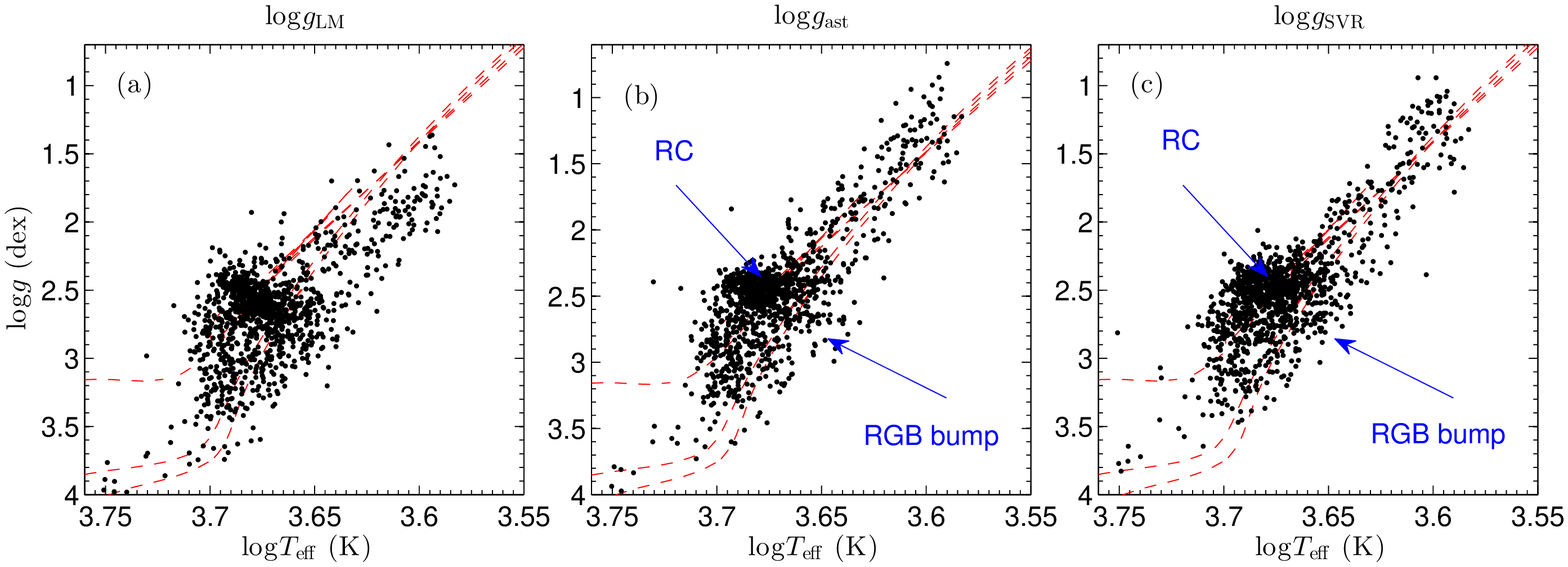}
\caption{The three panels show the distributions of the  {1,352} test dataset  {with \snrg$>20$} in \teff--\logg\ diagrams.  {The \logg\ in panel (a), (b), and (c) are from the LAMOST pipeline, the \emph{Kepler} asteroseismology, and this work, respectively.} The \teff\ values in the three panels are all from the LAMOST pipeline. The black dots are the test dataset and the red dashed lines show the isochrones with \feh$=-0.2$\,dex  {at the} age of 1, 5, and 10\,Gyr, from top to bottom. The red clump stars and the RGB bump stars in the second and third panels are indicated with texts and arrows}\label{fig:tefflogg}
\end{center}
\end{figure*}

 {In this section we use the test dataset with  {1,352} stars, which are not appeared in the training dataset but have seismic \logg\ and similar ranges of \teff\ and \feh\ as the training dataset, to investigate the performance. The re-calibrated seismic \logg\sei\ is used as the standard value to be compared with in the whole performance assessment.}

An intuitive assessment of the performance of the SVR model is to compare the SVR \logg\ (denoted as \logg\SVR) with other estimates in \teff--\logg\ diagram. Figure~\ref{fig:tefflogg} shows three \teff--\logg\ diagrams for the test dataset  {with \snrg$>20$}, the y-axes are the \logg\LM, \logg\sei, and derived \logg\SVR\ from panel (a) to (c), respectively. The x-axes, log\teffLM, in the three panels are all from the LAMOST pipeline. The red dashed lines show the isochrones \citep{marigo08} with \feh=-0.2\,dex at  {the age of} 1, 5, and 10\,Gyr, from  {top} to  {bottom}, respectively. 

\begin{figure}[htbp]
\begin{center}
\includegraphics[scale=0.5]{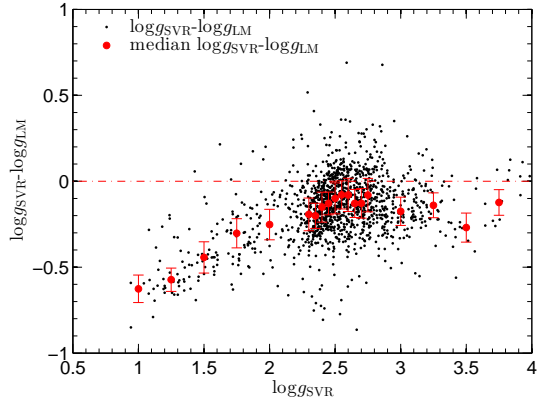}
\caption{The comparison between the SVR and the LAMOST \logg\ for the test dataset (the black dots). The red filled circles with the error bars are the median values and $1\sigma$ dispersions of the difference of \logg\ at each \logg\sei\ bin.}\label{fig:loggperform3}
\end{center}
\end{figure}

First, the largest difference between the \logg\LM\ and the other two is that the red giant branch stars (RGBs), which is located below the isochrones for \logg\LM( {panel (a)}). This can also be clearly seen in Figure~\ref{fig:loggperform3}, in which the difference of the \logg\SVR\ and \logg\LM\ is below the zero-point by 0.5\,dex for stars with \logg\SVR$<2$\,dex. Although it cannot be used to justify whether or not the LAMOST result is correct, the discrepancy between \logg\LM\ and the isochrone does bring significant systematic bias when one determine the distance by comparing \logg\LM\ and \teffLM\ with the synthetic isochrones. The seismic (panel (b)) and SVR (panel (c)) \logg, on the other hand, are consistent with the isochrones for the RGB stars.

Second, the most prominent feature in the \teff--\logg\ diagrams is the red clump/bump stars\footnote{ {Red clump stars are stars in the evolution phase of helium-core burning; Bump stars are the phase of first ascent RGB stars that a slight drop in the luminosities occurs when the extremely thin hydrogen shell is crossing to the discontinuous region\citep{cassisi97}}}, which concentrate at around \logg$\sim2.5$\,dex and log\teff$\sim3.68$. Compared with the seismic \logg, the LAMOST derived red clump stars (panel (a)) show an obviously tilted shape, which is probably an artificial effect due to the incompleteness and sparseness of the stellar library used in the current LAMOST pipeline. As the most accurate measurement of \logg, the seismic \logg\ shows the clear red bump located at $0.1\sim0.3$\,dex below the more concentrated and horizontally elongated red clump stars (\logg$\sim2.4$\,dex, $3.64<$log\teff$<3.7$) in panel (b).  {Encouragingly}, the SVR \logg\ shows similar features in panel (c): (1) the red clump is more concentrated than the LAMOST \logg\ and also shows a slightly elongated shape at $3.65<$log\teff$<3.7$; (2) the red bump stars can be barely discriminated just 0.3\,dex below the red clump at log\teff$\sim3.65$, although the dispersion is larger than the seismic \logg.  

 {Figure~\ref{fig:loggperform3} shows the difference between the SVR and LAMOST \logg. For the stars with \logg\SVR$>2.5$\,dex, \logg\SVR\ is slightly smaller than \logg\LM\ by about 0.1\,dex. For those with \logg\SVR$<2.5$\,dex, \logg\SVR\ is significantly smaller than \logg\LM\ by $\sim0.5$\,dex. This means that the LAMOST provided \logg\  may overestimate the \logg\ for the upper part of the RGB stars. }

\begin{figure}[htbp]
\begin{center}
\includegraphics[scale=0.5]{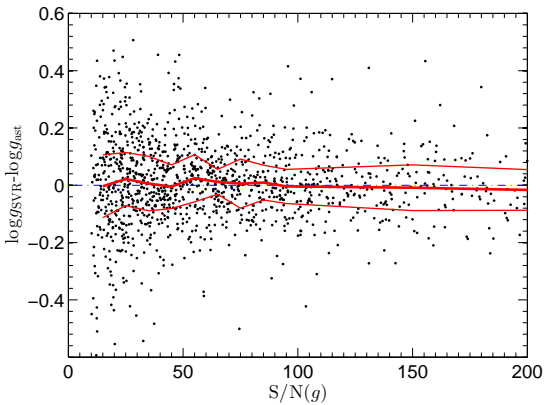}
\caption{The variation of the residual \logg\ of the test dataset with the signal-to-noise ratio at $g$ band of the corresponding LAMOST spectra (the black dots). The thick red line is the median value and the two thin red lines are the  {1}$\sigma$ of the dispersion. }\label{fig:loggperform2}
\end{center}
\end{figure}

 {Figure~\ref{fig:loggperform2} demonstrates the difference of the SVR with the seismic \logg\ as a function of the signal-to-noise ratio at $g$ band of the corresponding LAMOST spectra for the test dataset. The two thin red lines in Figure~\ref{fig:loggperform2} indicate the $1\sigma$ dispersion of the difference, which are flat at about $\pm$0.1\,dex when S/N($g$)$>20$. Although this uncertainty is still larger than the seismic measurement, it is significantly better than any other non-seismic estimation for the low-resolution spectra by a factor of 2-4. This explains why we can marginally distinguish the bump stars from the \teff--\logg\ diagram with SVR \logg\ in panel (c) of Figure~\ref{fig:tefflogg}.}

\begin{figure*}[ht]
\begin{center}
\includegraphics[scale=0.6]{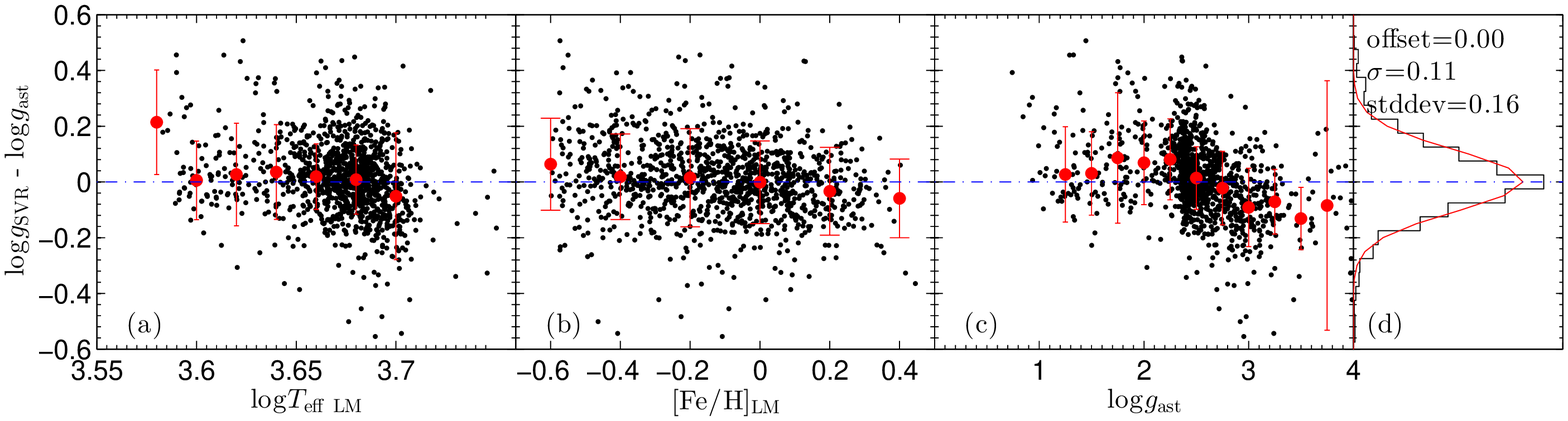}
\caption{The residual of the SVR \logg, \logg\SVR-\logg\sei\, for the test dataset  {with \snrg$>20$} as functions of the LAMOST log\teff\ and \feh\ are shown as the black dots in panels (a) and (b), respectively. Panel (c) shows the residual \logg\ as a function of the seismic \logg\ with the black dots. The red filled circles with error bars in the three panels indicate the medians and dispersions at  {various} positions. Panel (d) shows the distribution of the residual \logg\ with the black line. The red line in this panel is the  {best fit Gaussian, with parameters of offset=0 and $\sigma=0.11$\,dex}, to the residual. The standard deviation of the residual is  {0.16\,dex}.}\label{fig:loggperform}
\end{center}
\end{figure*}

 {Figure~\ref{fig:loggperform} shows the performance of the SVR \logg\, compared with the corresponding seismic \logg\ using the test dataset  {with \snrg$>20$}. First, the comparisons between the residual \logg\ (defined as \logg\SVR-\logg\sei) does not show any strong correlation with the effective temperature, as shown in panels (a). Second, it shows very weak anti-correlation with \feh, i.e., the SVR \logg\ is very slightly overestimated for stars with \feh$\sim-0.6$\,dex and underestimated for those with \feh$\sim+0.2$\,dex. This systematic bias is below 0.05\,dex, within the dispersion of the residual of \logg\SVR. Third, in panel (c), for stars with \logg\sei$>3$\,dex, the SVR \logg\ is underestimated by $\sim$0.1\,dex. Then the residual of \logg\SVR\ changes from negative to positive when \logg\sei\ changes from 3 to 2.25\,dex. Consider that the dispersion of the residual in this range is slightly larger than 0.1\,dex, this systematics is not significant, although it is not negligible. For those \logg\sei$<2.25$\,dex, a slight overestimation of less than 0.05\,dex in \logg\SVR\ is found. This inverse S-shape in the residual \logg\SVR\ vs. \logg\sei\ will be further discussed in section~\ref{sect:evophases}.
Finally, the distribution of the residual gives the overall uncertainty of the measurement in panel (d). The standard deviation for the residual is 0.16\,dex. We also fit the distribution of the residual with a Gaussian to give an alternative estimation of the dispersion. The best fit Gaussian (the red line) is centered at 0, meaning that there is no overall systematic bias in the SVR estimation, and $\sigma$ is 0.11\,dex.}

\subsection{Apply to LAMOST data}\label{sect:apply}
\begin{figure}
\begin{center}
\includegraphics[scale=0.6]{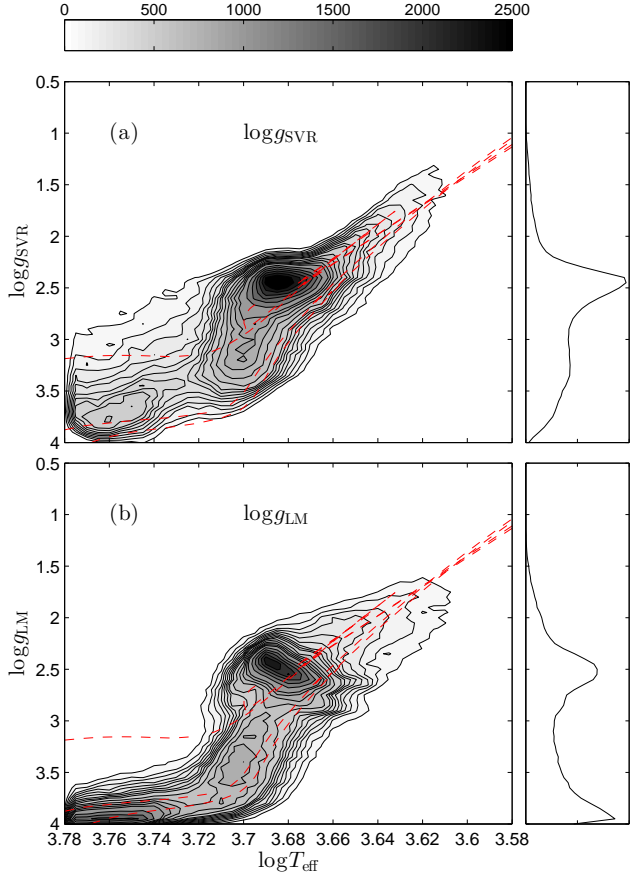}
\caption{Panel (a) shows the distribution of the LAMOST metal-rich giant stars in log\teff\ vs. \logg\ plane. The \teff\ is from the LAMOST pipeline, while the \logg\ is estimated from the seismic-based SVR. The  {contours indicate} the density. Panel (b) shows the same distribution for the same dataset but with the LAMOST \logg. The red dashed lines show the isochrones with \feh$=-0.2$\,dex  {at the} age of 1, 5, and 10\,Gyr, from top to bottom.  {The distribution of the two estimated \logg\ are shown in the right-side panels as the black lines. }}\label{fig:tefflogglamost}
\end{center}
\end{figure}

We apply the SVR \logg\ estimator to  {356,932} selected LAMOST DR2  {stars} with \logg\LM$<4$\,dex, \feh\LM$>-0.6$\,dex, and reliably measured equivalent widths for all 23 line indices. The final result is shown in panel (a) of Figure~\ref{fig:tefflogglamost}. As a comparison, panel (b) shows a similar  {plot} with the LAMOST derived \logg. Compared with the theoretical isochrones, the underestimation in the LAMOST \logg\ of the RGB stars has been corrected in the  {results of this work}. And the shape around the red clump stars now seems normal.  {However, a small fraction of stars with log\teff$>3.72$ are significantly underestimated in \logg\ because the poor quality of the corresponding spectra for these stars.}

\section{Discussions}\label{sect:discussions}

\subsection{The systematic bias in seismic parameter determinations}\label{sect:seismicbias}

 {The SVR \logg\ is automatically calibrated to the seismic \logg, since it uses the latter as the standard value in the training dataset. This also means that all systematics occur in seismic \logg\ will also come into SVR \logg. Therefore, it is important that the users, particularly those who are not working in asteroseismology, of the SVR \logg\ should be aware of the performance of the seismic \logg. Quite a lots of works have been done to investigate the performance of the seismic scaling relations. Here we just briefly review a few of them related to \logg. Detailed review is referred to~\citet{chaplin13}.}

 {According to Equation~(\ref{eq:scleqn}), the accuracy of the surface gravity depends on a few factors, the frequency of maximum power $\nu_{max}$, the effective temperature \teff, and the corresponding solar parameters. \citet{gai11} investigated the grid-based method to derive stellar properties, including \logg, from seismic parameters combined with effective temperature and metallicity. They found that the uncertainty (in the term of FWHM of the distribution of the fractional deviation of the true value) is smaller for star with large \logg\ than for that with small \logg. In all cases, the uncertainty is within 5\% of the true \logg. \citet{huber12} combined the asteroseismology, interferometry, and parallax of the stars and found that the uncertainty of the seismic scaling relations are in agreement with other measurement; and the uncertainties do not depend the evolution stage of the stars. However, \citet{hekker13} inferred that the seismic parameter determinations give small systematic bias in \logg\ for both red clump and RGB stars, although the extents of the systematics is within 0.01\,dex. More investigations about the various evolution phases of stars are stated in section~\ref{sect:evophases}. }

 {All these systematic bias do affect the SVR estimates for the LAMOST data, since the SVR treats the seismic \logg\ as the standard. Therefore, the users who intend to apply the \logg\SVR\ in any study of the structure and evolution of the Milky Way should very carefully validate whether these systematic bias affect their results.}
 
 {It is also worthy to note that the limit of the techniques and observations in asteroseismology may bias the sampling of the giant stars at different \logg. For the giant stars with larger \logg, because their oscillation amplitude is very small \citep{huber11} and thus hard to be detected, many this kind of stars in the \emph{Kepler} field are lack asteroseismic \logg. On the other hand, the giant stars with smaller \logg\ are also under-sampled because their oscillation frequency is too low and cannot be reliably measured during the 4-year observations with \emph{Kepler}. Therefore, the training dataset is lack data at both ends of the distribution of the \logg\sei. This may lead to some systematics in the SVR method because the imbalance of the training dataset. Fortunately, it seems that the training dataset contains sufficient stars with either very small or very large \logg\sei\ and the systematic bias shown in panel (c) of Figure~\ref{fig:loggperform} is acceptable, compared to the uncertainty of $\sim0.1$\,dex.}

\begin{figure}[htbp]
\centering
\includegraphics[scale=0.5]{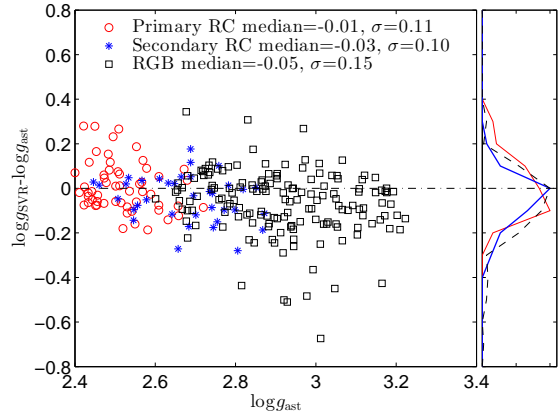}
\caption{ {The figure shows the residual \logg\ as a function of the seismic \logg\ for the primary (red circles), secondary red clump stars (asterisks), and RGB stars (black circles). The right-hand panel shows the distribution of the three types of stars with the same colors as the main plot.}}\label{fig:evostage}
\end{figure}

\subsection{Effect of evolution phases}\label{sect:evophases}
 \citet{pinsonneault14} found that the spectroscopic and seismic \logg\ are systematically different for the giant stars in different evolution stage, e.g., the primary red clump stars, secondary red clump stars\footnote{ {The primary red clump stars are the He-core burning giant stars with electronic-degenerate core, while the secondary clump stars are massive enough to have non-degenerate He-cores. In general, the primary clump stars are low-mass and hence older. However, the secondary clump stars are younger and the present stellar model predicts that they are around 1\,Gyr old~\citep{girardi99}.}}, and RGB stars. Although the training dataset of the SVR model uses seismic \logg, the input data are based on the spectra. Therefore, it may also have similar systematic bias as shown in figure 3 of~\citet{pinsonneault14}. We then tag the primary, secondary red clump stars and RGB stars in the test dataset by cross-identifying it with the data from~\citet{stello13}, in which the evolution phases are distinguished by the period spacings of the dipole mode. {Figure~\ref{fig:evostage} shows that \logg\SVR\ is about 0.05\,dex lower than \logg\sei\ for the RGB stars.} Although almost no overall systematics is found for primary red clump stars, a few of them are overestimated by about 0.2\,dex (see the left side of figure~\ref{fig:evostage}). For the secondary red clump stars, the median residual slightly shift to -0.03\,dex. These can explain why in the panel (c) of Figure~\ref{fig:loggperform} the SVR \logg\ is systematically underestimated for stars with \logg\sei$>3$\,dex, which are mostly RGB stars, and then slightly overestimated for \logg\sei$<2.25$\,dex. Moreover, the dispersion of the residual between \logg\SVR\ and \logg\sei is larger in RGB stars than in red clump stars. {It seems that the effect of evolution phases in LAMOST is different with that in the APOGEE data \citep{pinsonneault14}. In future works, more investigations are necessary to address the reason why the different evolution stages show different bias in \logg\ estimates and to cross-calibrate the parameters between LAMOST and APOGEE data.}

\subsection{Metallicity}\label{sect:metallicity}
\begin{figure}[htbp]
\centering
\includegraphics[scale=0.5]{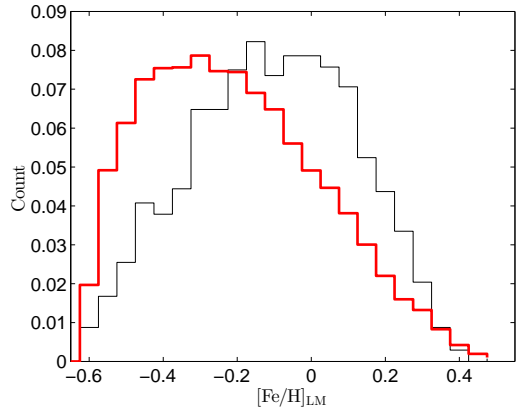}
\caption{ {The metallicity distribution of the training dataset (black line) and the LAMOST DR2 data (red line). The y-axis is the count of stars normalized by 1}}\label{fig:mdf}
\end{figure}

Currently, the seismic \logg\ is limited to the metal-rich stars. This is because the empirical scaling relation used in the asteroseismology measurement is based on the solar-type stars. \citet{epstein14a} found that the seismic mass estimation for the halo and thick disk stars is significantly higher than the expectations. Although the stellar radius estimation is more precise than the stellar mass \citep{gai11}, it may also be affected by the same systematic bias. Therefore, we only apply the seismic-trained SVR model to the metal-rich giant stars (\feh$>-0.6$\,dex) to avoid the probably systematic shift for the metal-poor stars. 

 {It is also worthy to compare the metallicity distribution function (MDF) between the training dataset and the selected LAMOST data to ensure that the training dataset covers the all range of the metallicity of the LAMOST data. Figure~\ref{fig:mdf} shows the two MDFs (the gray and red lines stand for the MDF of the training dataset and the LAMOST data, respectively). Because both the two samples are cut at -0.6\,dex, the two MDFs are truncated at this value.  Although the peaks of the two distributions are different by about 0.2\,dex, they approximately covers the same range of the metallicity, except a very small fraction of data at about 0.4\,dex. Therefore, the scaling relation of the seismology used in the surface gravity estimation for the training dataset is also suitable for the selected LAMOST data.} 

 {Moreover, the panel (b) of figure~\ref{fig:loggperform} shows that \logg\SVR\ does not strongly correlated with the metallicity. Therefore, the slight difference of the MDFs shown in Figure~\ref{fig:mdf} would not lead to significant systematics in the derived \logg\ for the LAMOST data.}

\subsection{Compare with other works}\label{sect:compare}
 {\citet{meszaros13} and~\citet{holtzman15} calibrated the surface gravity of the APOGEE~\citep{ahn14} data to seismic values with simple functions. Compared to the technique employed in this work, their methods are direct and simple. However, the measurement errors in their methods are contributed twice from two different processes, one is from the estimation of the initial \logg\ based on synthetic library and the other is from the calibration after \logg\ estimation. Therefore, the uncertainty of the calibrated \logg\ may be intrinsically larger. In our method, the seismic \logg\ is directly used as the standard values in training process, and therefore, the measurement error is only produced once during the determination of the \logg. The derived \logg\ from SVR model is naturally calibrated with the seismic \logg. In other word, the SVR model merges the two steps of determination and calibration of \logg\ into one single step and hence avoid additional error contribution.}

\subsection{Degeneracies between \logg, \teff, and \feh}

{Figures~\ref{fig:EWlogg} --\ref{fig:EWfeh} show that the stellar parameters, \logg, \teff, and \feh, do not precisely orthogonal in the spectral lines. In some of the lines, the stellar parameters even change toward similar directions, producing some extents of degeneracies. It is quite difficult to completely break the degeneracies in a stellar parameterization technique, particularly in the algorithm like SVR, which can only determine one parameter at a time. Indeed, figure~\ref{fig:loggperform} does show that the residuals of \logg\SVR\ bias from zero by at most $\sim0.05$\,dex in \teff\ and \feh. However, considering that the \logg\SVR\ estimated in this work will be mainly used in the large sample statistics with tens or even hundreds of thousands of stars in the Galaxy, these weak biases may not be significantly affect the results.}

{It is also noted that the SVR \logg\ may also change the estimation of \teff\ and \feh. Figures~\ref{fig:EWlogg} --\ref{fig:EWfeh} show that some spectral lines are sensitive to all of the three parameters. This means that when we estimate \teff\ and \feh\ with the adopted SVR \logg\ values, the derived \teff\ and \feh\ may quite different with the current values from the LAMOST pipeline. The difference can be quantified by re-estimating the \teff\ and \feh\ in the LAMOST pipeline with the fixed SVR \logg\ value. This, however, is beyond the scope of this paper and should be investigated in future works.}

\subsection{Benefits from accurate \logg}\label{sect:benefits}

 {We can significantly improve the accuracy of the distance estimation from accurate \logg\ estimates.} Since the stellar luminosity follows $L\propto R^2$ and surface gravity follows $g\propto R^{-2}$, we have $L\propto g^{-1}$. Consequently, the absolute magnitude $M$ is proportional to ${\rm log}L\sim2.5{\rm log}g$. This means when the uncertainty of \logg\ is 0.1\,dex, the uncertainty of the absolute magnitude turns out to be 0.25\,mag, which is equivalent with about 12\% in distance\footnote{{Note that the accuracy of the distance does not take into account the uncertainties contributed by photometries, e.g., the coarse dereddening, less accurate magnitude etc.}}. This is a factor of $\sim$2 better than the accuracy of distance derived from the non-seismic based spectroscopic \logg~\citep{carlin15}. 


\subsection{The dwarf stars}\label{sect:limits}
The known \emph{Kepler} seismic \logg\ are mostly for the giant stars \citep{huber14}. Hence, we lack dwarf star samples as the training dataset. As a consequence, we only apply the method to giant stars and not expand it to dwarf stars. The LAMOST \logg\ is sufficiently accurate for the separation of the giant and dwarf stars. Therefore, we use it to select the giant stars first and let the seismic-trained SVR model to predict more accurate \logg\ for the selected giant stars. In the future, the coming PLATO mission may provide another tens of thousands asteroseismic measurement over the entire Hertzsprung-Russell diagram \citep{rauer14}. This will help to enrich the dwarf training dataset and expand this work to the dwarf stars.

\section{Conclusions}\label{sect:conclusions}
 {Although we use the LAMOST spectra as the training dataset, it does not limit the application of the SVR model only to the LAMOST data. In general, given another spectroscopic survey data without seismic measurement, we can firstly calibrate the equivalent widths of the line indices to align with the LAMOST and then apply the model to the new data. Indeed, we test it using a set of MMT/Hectospec observed giant stars, in which more than 100 common objects with the LAMOST are found. The accuracy of the \logg\ estimates for these samples is roughly at the same level as in this work (more details will be given in Liu et al. in preparation).}

In summary, although not all the LAMOST data have asteroseismic observations, we can use a small subset with the \emph{Kepler} seismic \logg\ as the training dataset to estimate \logg\ for other LAMOST data with a support vector regression model. The approach can reach to an accuracy as high as 0.1\,dex when the signal-to-noise ratio of the spectra is higher than 20. This improves the current \logg\ estimated from the LAMOST pipeline by at least a factor of 2. This significant improvement will be very useful in the following studies: 1) it allows us to better estimate the distance of the giant stars  {with accuracy of about 12\%}; and 2) it enables to separate the primary and secondary red clump stars from \logg, providing good samples to trace the stellar populations with different ages.
 
\acknowledgments
 {We thank the anonymous referee for his/her useful comments.} We also thank Ren\'e Andrae, Coryn Bailer-Jones, and Shude Mao for the helpful comments and discussions. This work is supported by the Strategic Priority Research Program
"The Emergence of Cosmological Structures" of the Chinese Academy of Sciences, Grant No. XDB09000000 and the National Key Basic
Research Program of China 2014CB845700. CL acknowledges the National Science Foundation of China (NSFC) under grants 11373032, 11333003, and U1231119. MF acknowledge the NSFC under grant 11203081. YW acknowledge the NSFC under grant 11403056. Guoshoujing Telescope (the Large Sky Area Multi-Object Fiber Spectroscopic Telescope LAMOST) is a National Major Scientific Project built by the Chinese Academy of Sciences. Funding for the project has been provided by the National Development and Reform Commission. LAMOST is operated and managed by the National Astronomical Observatories, Chinese Academy of Sciences.

{\it Facilities:} \facility{LAMOST}.



\end{document}